\documentstyle[prb,aps,psfig,twocolumn]{revtex}
\begin{document}
\draft

\setlength{\baselineskip}{1.0em}
\setlength{\parskip}{0ex}
\addtolength{\textwidth}{3cm}
\addtolength{\textheight}{2cm}
\hoffset-4mm

\title{Ferromagnetism and Temperature-Driven Reorientation Transition in Thin 
        Itinerant-Electron Films}
\author{T. Herrmann, M. Potthoff, and W. Nolting}
\address{Humboldt-Universit\"at zu Berlin, Institut f\"ur Physik,               
           Invalidenstr.\ 110, 10115 Berlin, Germany}

\maketitle
%
\begin{abstract}
The temperature-driven reorientation transition which, up to now, has been 
studied by use of Heisenberg-type models only, is investigated 
within an itinerant-electron model. We consider the Hubbard model for a 
thin fcc(100) film together with  the dipole
interaction and a layer-dependent anisotropy field. 
The isotropic part of the model is treated
by use of a generalization of the spectral-density approach to
the film geometry.  The magnetic properties of the film 
are investigated as a function of
temperature and film thickness and are analyzed in detail with help of the
spin- and layer-dependent quasiparticle density of states. 
By calculating the temperature dependence of
the second-order anisotropy constants we find that both types of reorientation
transitions, from out-of-plane to in-plane (``Fe-type'') 
and from in-plane to out-of-plane (``Ni-type'') magnetization are
possible within our model. In the latter case the
inclusion of a positive volume anisotropy is vital.  The reorientation
transition is mediated by a strong reduction of the surface magnetization with
respect to the inner layers as a function of temperature and is found to depend
significantly on the total band occupation.

\end{abstract}
%
%
%
%
%
%
%
%
\renewcommand{\baselinestretch}{1.1}
\Large\normalsize
\section{Introduction}
The large variety of novel and interesting
phenomena of thin-film magnetism results 
very much from the fact that the magnetic
anisotropy, which determines the easy axis of magnetization, can be 
one or two orders
of magnitude larger than in the corresponding bulk systems\cite{Nee54}.
The reorientation transition (RT) of the direction of magnetization  in
thin ferromagnetic films describes the change of the easy axis 
by variation of  the film thickness or temperature and
has been widely studied both experimentally 
\cite{PKH90,PBH92,QPB93,ASB90,AB92,All94,FKEG94,FB96,LSK+97,%
SB94,Bab96,BFW97,FPA+97b,FPA+97a} and 
theoretically 
\cite{TBF76,Bru89,CEM94,LMH97,WWF93,DKS94,LH96,HBW+97,SNF97,%
CC66,LLS69,PP90,JB90,JB96,Jen97,MU95a,MU95b,HU97,MWDP96,MDW98}. 

An instructive phenomenological picture for the understanding of the RT is
obtained by expanding
the free
energy $F$ of the system in powers of $\cos\theta_M$, where  $\theta_M$ is 
the angle between the direction of magnetization and the surface normal. 
Neglecting azimuthal anisotropy and exploiting time inversion symmetry
yields:
\begin{equation}
  F (\theta_M)= F_0 - K_2 \cos^2\theta_M - K_4 \cos^4\theta_M-\dots \,\,.
\end{equation}
The anisotropy coefficients of second ($K_2$) and fourth ($K_4$) order
depend on the thickness $d$  of the film as well as on the temperature $T$. 

Away from the transition point usually $K_2\gg K_4$ holds,  and, therefore, 
the direction of
magnetization is determined by the sign of $K_2$ 
($K_2>0$: out-of-plane magnetization; $K_2<0$: in-plane
magnetization).
On this basis the concept  of anisotropy flow \cite{MK96,Bab96} 
immediately tells us that 
the RT is caused by a sign change of $K_2$   while
the sign of $K_4$ mainly determines whether the transition 
is continuous ($K_4<0$) or
step-like ($K_4>0$). In the case of a 
continuous transition $K_4$ also gives the
width of the  transition region.

From the microscopic point of view we know  that the magnetic anisotropy is
exclusively caused by two effects, the dipole interaction between the magnetic
moments  in the sample and the spin-orbit coupling:
$K_i=K_{i,\text{so}}+K_{i,\text{dip}}$. 
While the dipole interaction
always favors in-plane magnetization 
($K_{2,\text{dip}}<0$) due to minimization
of stray fields, the spin-orbit interaction can lead to both, in-plane and
out-of-plane magnetization depending sensitively on the electronic structure of
the underlying sample. 
The spin-orbit anisotropy  is caused by the broken symmetry\cite{Nee54} 
at the film surface and the substrate-film interface
as well as by possible strain\cite{HBW+97,SNF97} in the volume of the film. 
It is worth to stress that a strong positive spin-orbit induced anisotropy
alone opens up the possibility of an out-of-plane magnetized thin
film. The RT must be seen as a  competition between spin-orbit and dipole
anisotropy.

In many thin-film systems both thickness- and temperature-driven RTs 
are observed.
Although it is clear by inspection of the corresponding phase diagrams
\cite{PKH90,FPA+97b} that both types of transitions are closely related to
each other, different theoretical concepts are needed to explain their
physical origin.  

The thickness-driven RT is rather well 
understood in terms of a phenomenological separation 
of the spin-orbit induced anisotropy constant $K_{i,\text{so}}$ 
into a surface term $K_{i,\text{so}}^S$ and  a 
volume contribution $K_{i,\text{so}}^V$ by the ansatz 
$K_{i,\text{so}}=K_{i,\text{so}}^V+
2 K_{i,\text{so}}^S/d$. 
Experimentally, this separation seems to provide a rather consistent     
picture\cite{All94,FKEG94,SB94,Bab96} despite the fact that in some samples 
additional structural transitions are present\cite{FB96,LSK+97} which clearly 
restrict its  validity. 
On the theoretical side, basically 
two different schemes 
for the calculation of $T=0$ magnetic anisotropy constants
have been developed, semi-empirical
tight-binding theories\cite{TBF76,Bru89,CEM94,LMH97} and 
spin-polarized ab initio total-energy calculations
\cite{WWF93,DKS94,LH96,HBW+97,SNF97}.
In both approaches the spin-orbit coupling 
is introduced either self-consistently
or as a final perturbation. However, these investigations still remain to be 
a delicate problem
because of the very small energy differences
involved.

Neglecting the large variety of different 
samples, substrates, growth conditions,
etc. it is useful for the understanding of the RT 
to concentrate on two somewhat
idealized prototype systems both showing a thickness-
as well as a temperature-driven RT.

The ``Fe-type'' systems \cite{PKH90,PBH92,QPB93,ASB90,AB92,All94} 
are characterized by a large positive
surface anisotropy constant $K_2^S$ together with a negative volume anisotropy
$K_2^V$ due to dipole interaction. This leads to out-of-plane
magnetization for very thin  films. For increasing film thickness  
the magnetization switches to an in-plane direction  because 
the volume contribution  becomes dominating\cite{ASB90,AB92,All94}. 
As a function of increasing temperature a RT
from out-of-plane to in-plane magnetization is found for certain
thicknesses \cite{PKH90,PBH92,QPB93}.

In the ``Ni-type'' systems \cite{SB94,Bab96,BFW97,FPA+97b,FPA+97a},
the situation is different. Here the volume contribution
$K_2^V$ is positive due to fct lattice distortion \cite{Bab96,HBW+97}, thereby 
favoring out-of-plane magnetization, while the surface term $K_2^S$ is
negative. For very thin films the surface contribution dominates leading 
to  in-plane
magnetization. At a critical thickness, however, the positive volume anisotropy
forces the system to be magnetized in out-of-plane direction 
\cite{SB94,Bab96,BFW97}, 
until at a second
critical thickness the magnetization switches to an in-plane position again
caused by structural relaxation effects. 
Here a so-called anomalous 
temperature-driven RT from in-plane to out-of-plane magnetization
was found recently by Farle et~al.\cite{FPA+97b,FPA+97a}.

In this article we will focus on the temperature-driven RT which cannot
be understood by 
means of the separation into surface and volume contribution alone.
Here the coefficients $K_i^S$ and $K_i^V$ need to be determined
for each temperature separately. Experimentally, this has been done in great
detail for the second-order anisotropy of Ni/Cu(100)\cite{FPA+97b}. 
The results clearly confirm
the existence and position of the RT, but, 
on the other hand, do not lead to any
microscopic understanding of its origin.

To obtain more information on the temperature-driven RT theoretical 
investigations on 
simplified model systems have proven to be fruitful. Despite the fact that in
the underlying transition-metal samples the spontaneous magnetization is caused
by the itinerant, strongly correlated 3d-electrons, up to now  Heisenberg-type
models have been considered exclusively
\cite{CC66,LLS69,PP90,JB90,JB96,Jen97,MU95a,MU95b,HU97,MWDP96,MDW98}. 
The magnetic anisotropy has been taken into account by incorporating the 
dipole interaction and an uniaxial single-ion anisotropy to model the 
spin-orbit-induced  anisotropy.
Using appropriate  
$T=0$ second-order
anisotropy constants as  input parameters, both types of RTs have been observed
within the framework of a self-consistent mean field approximation 
\cite{HU97,Jen97}
as well as  by first-order perturbation theory for the free energy \cite{JB96}. 
A continuous RT has been found for $d\ge 3$ layers\cite{HU97,Jen97} 
taking place over a rather small temperature range. 
Step-like transitions occur  as an exception for special parameter
constellations only. The RT is attributed to the strong reduction 
of the surface-layer magnetization relative to the inner layers for 
increasing temperature
leading to a diminishing influence of the surface anisotropy.

Since the itinerant nature of the magnetic moments is ignored completely in
these calculations, 
it is interesting to compare these results with calculations done within
itinerant-electron systems.
The present work  employs a similar concept but in the framework of the 
single-band Hubbard model\cite{Hub63} which we believe is a more reasonable 
starting point for the description of temperature-dependent 
electronic structure of thin transition-metal films.

The paper is organized in the following way: 
In the next section we define our model
Hamilton operator. In Sec.~\ref{sec_f} we will focus on the derivation of the
free energy and the second-order anisotropy constants by use of a
perturbational approach.  The isotropic part of the Hamilton operator is
treated in Sec.~\ref{sec_sda}. Here we present a generalization of a 
self-consistent spectral-density approach to the film geometry. 
In Sec.~\ref{sec_res} we will show and analyze the results of the numerical 
evaluations  and discuss the possibility of a temperature-driven 
RT within our model system. 
We will end with a short conclusion in Sec.~\ref{sec_con}.

%
%
%
%
%
%
%
%
%
\section{Definition of the Hamilton operator}

The description of the film geometry requires some care. Each lattice vector of 
the film is decomposed into two parts
\begin{equation}
    {\bf R}_{i\alpha}={\bf R}_i+{\bf r}_\alpha.
\end{equation}      
${\bf R}_i$ denotes a lattice vector of the underlying two-dimensional
Bravais lattice with $N$ sites. 
To each lattice point a $d$-atom basis ${\bf r}_\alpha$ ($\alpha=1,\dots,d$) 
is associated referring to the $d$ layers of
the film. 
The same labeling, of course, also applies  for all other 
quantities related to the film geometry.
Within each layer we assume translational invariance. Then a Fourier
transformation with respect to the two-dimensional 
Bravais lattice can be applied.

The considered model Hamiltonian consists of three parts:
\begin{equation}
H=H_0+H_{\text{dip}}+H_{\text{so}}.\label{ham_op}
\end{equation}
$H_0$ denotes the single-band Hubbard model 
\begin{equation} \label{hub_op}
        H_0=\sum_{i,j,\alpha,\beta,\sigma}
	(T_{ij}^{\alpha\beta}-\mu\delta_{ij}^{\alpha\beta})
	c_{i\alpha\sigma}^{\dagger}c_{j\beta\sigma}+
	\frac{U}{2}\sum_{i,\alpha,\sigma}n_{i\alpha\sigma} n_{i\alpha-\sigma},
\end{equation}
where $c_{i\alpha\sigma}$ ($c_{i\alpha\sigma}^{\dagger}$) stands for 
the annihilation
(creation) operator of an electron with spin $\sigma$ at the lattice site
${\bf R}_{i\alpha}$, 
$n_{i\alpha}=c_{i\alpha\sigma}^{\dagger}c_{i\alpha\sigma}$ 
is the number operator and $T_{ij}^{\alpha\beta}$ is the hopping-matrix element
between the lattice sites ${\bf R}_{i\alpha}$ and ${\bf R}_{j\beta}$. The
hopping-matrix element between nearest neighbor sites is set to $-t$.
$U$ denotes the on-site Coulomb matrix element, and $\mu$ 
is the chemical potential.

The second term $H_{\text{dip}}$ 
describes the dipole interaction between the
magnetic moments on different lattice sites:
\begin{equation} \label{dip_op}
H_{\text{dip}}=\frac{\omega_0}{2}
\sum_{i,j,\alpha,\beta}^{i,\alpha\neq j,\beta}
\frac{1}{(\frac{r_{ij}^{\alpha\beta}}{a})^3}
\Big[ \bbox{\sigma}_{i\alpha} \bbox{\sigma}_{j\beta} 
-3(\bbox{\sigma}_{i\alpha} 
\hat {\bf u}_{ij}^{\alpha\beta})(\bbox{\sigma}_{j\beta} 
\hat {\bf u}_{ij}^{\alpha\beta})\Big].
\end{equation}

Here 
$r_{ij}^{\alpha\beta}$ is the distance between ${\bf R}_{i\alpha}$ and
${\bf R}_{j\beta}$, and the unit vector 
$\hat {\bf u}_{ij}^{\alpha\beta}$ is given by
$\hat {\bf u}_{ij}^{\alpha\beta}=
({\bf R}_{i\alpha}-{\bf R}_{j\beta})/r_{ij}^{\alpha\beta}$.  
$\omega_0=\mu_0\mu_B^2 z_{uc}/(4\pi a^3)$ denotes the
strength of the dipole interaction, $a$ the lattice constant and $z_{uc}$ the
number of atoms in the corresponding bulk cubic unit cell.
$\bbox{\sigma}_{i\alpha}=\sum_{\tau\tau'}
       c_{i\alpha\tau}^{\dagger}\bbox{\sigma}_{\tau\tau'}
       c_{i\alpha\tau'}$
are the spin operators constructed by  
the Pauli spin matrices $\bbox{\sigma}$. 
The expectation value of the spin
operator yields the (dimensionless) magnetization vector
\begin{equation}
{\bf m}_\alpha=\langle \bbox{\sigma}_{i\alpha}\rangle.
\end{equation} 
The magnetization only depends  on the layer index $\alpha$ because of 
the assumed translational invariance within the layers.

In our approach the uniaxial anisotropy caused by the spin-orbit coupling is
taken into account phenomenologically by an effective 
layer-dependent anisotropy field coupled 
to the spin operator:
\begin{equation} \label{so_op}
        H_{\text{so}}=-\sum_{i}  
        {\bf B}_{\alpha}^{\text{(so)}}
        \bbox{\sigma}_{i\alpha}.
\end{equation}
The effective field ${\bf B}_{\alpha}^{\text{(so)}}$ 
is chosen to be 
parallel to the
film normal $\hat{\bf n}$. To ensure the right symmetry we set:
\begin{equation}\label{anis_field}
    {\bf B}_{\alpha}^{\text{(so)}}=
    \beta_{\alpha}({\bf m}_{\alpha} \hat{\bf n})\cdot \hat{\bf n}.
\end{equation}
This corresponds to a mean-field treatment of the spin-orbit-induced
anisotropy\cite{JB90}. 
The strengths of the anisotropy 
fields $\beta_\alpha$ enter as additional parameters and 
have to be fixed later.

%
%
%
%
%
%
%
%
%
%
%

\section{Calculation of the magnetic anisotropy constants}
\label{sec_f}

The direction of the magnetization is determined 
by the minimal free energy $F$.
The anisotropic contributions $H_{\text{so}}$ 
and $H_{\text{dip}}$ to the Hamilton
operator (\ref{ham_op})  can be considered as a small perturbation to the
isotropic Hubbard model  ($\beta_\alpha,\,\,\omega_0\ll t,\,\,U$). 
Then we
can apply a thermodynamic perturbation 
expansion\cite{CC66,LLS69} of the free energy $F$ up to
linear order with respect to $\beta_\alpha,\,\,\omega_0$:
\begin{equation} \label{f_pert}
F(T)=F_0(T) + \langle H_{\text{so}}\rangle_0/(Nd) +
              \langle H_{\text{dip}}\rangle_0/(Nd).
\end{equation}
Here $\langle \dots \rangle_0$ denotes the expectation value taken 
within the unperturbated Hubbard model $H_0$.
On the same footing we use a mean-field decoupling for the two-particle
expectation values contained in 
$\langle H_{\text{dip}}\rangle_0$. 
Because $F$ is calculated to linear order in the anisotropy
contributions, only the lowest, i.e. second-order anisotropy constants are
considered, and a possible canted phase is neglected. 
The ratio $K_4/K_2$ and thus the  width of the transition
region have been found to be very small 
for reasonable strengths of the anisotropy
contributions \cite{HU97,Jen97}.

Within our approach  it is, therefore,  sufficient to consider the 
free-energy difference $K_2(T)$ between in-plane and out-of-plane
magnetization: 
\begin{equation}
   K_2(T)=F(T,\theta_M=\frac{\pi}{2})-F(T,\theta_M=0).
\end{equation}
$K_2<0$ and $K_2>0$ indicate in-plane and
out-of-plane magnetization respectively.
Hence, the reorientation temperature $T_{R}$ is  given by the condition
$K_2(T_R)=0$.
Evaluation of $K_2(T)$ yields:
\begin{eqnarray}
   K_2(T)&=&K_{2,\text{so}}+
   K_{2,\text{dip}}\nonumber\\
    &=&\frac{1}{d}\sum_\alpha \beta_{\alpha}m_{\alpha}^2 -\frac{3\omega_0}{2}
   \frac{1}{d}\sum_{\alpha\beta}m_\alpha
   m_\beta\Phi_{|\alpha-\beta|}.\label{f_anis_end}
\end{eqnarray}
The constants $\Phi_\alpha$ contain the effective dipole interaction between the
layers and can be calculated separately:
\begin{equation}
   \Phi_{|\alpha-\beta|}=\frac{1}{z_{uc}}\frac{1}{N}\sum_{i,j}^{i\ne j}
   \frac{1}{(\frac{r_{ij}^{\alpha\beta}}{a})^3} 
   \left[\cos^2\theta_{i\alpha j\beta}^{(90)}-\cos^2\theta_{i\alpha
   j\beta}^{(0)}\right].
\end{equation}
$\theta_{i\alpha j\beta}^{(\theta_M)}$ is the angle between 
$\hat {\bf u}_{ij}^{\alpha\beta}$ and the direction of the magnetization.
The $\Phi_\alpha$ only depend on the film geometry. 
For thick films  $K_{2,\text{dip}}$
reduces to its continuum value $\frac{1}{2}\mu_0 M^2$ where $M$ is the
magnetization per atom. To calculate $K_2(T)$  the
temperature- and layer-dependent magnetizations $m_\alpha$ of 
the  Hubbard film are needed.

%
%
%
%
%
%
%
%
%
%
%
\section{Spectral-density approach to the Hubbard film}\label{sec_sda}
In this section we will focus on the evaluation of the Hubbard film
$H_0$. 
Ferromagnetism in the Hubbard model is surely a strong-coupling
phenomenon. The
existence of ferromagnetic solutions was recently proven  in the limit of
infinite dimensions by quantum Monte-Carlo calculations\cite{Ulm98,VBH+97}. 
Ferromagnetism is favored by a strongly asymmetric Bloch
density of states (BDOS) and by a singularity at the upper band edge
as it is found, e. g., for the fcc lattice.

The Hubbard model constitutes a highly non-trivial many-body problem
even for a periodic infinitely extended lattice. Even more complications are
introduced when the reduction of translational symmetry has to be taken into
account additionally.
One of the easiest possible approximations to treat a Hubbard film is a
Hartree-Fock decoupling, which has been applied previously \cite{GM94,PM95}. 
Hartree-Fock theory, however, is necessarily restricted to the weak-coupling
regime and is known to overestimate the possibility of ferromagnetic
order drastically. Neglecting electron-correlation effects altogether 
leads to qualitatively wrong results especially for intermediate 
and strong Coulomb interaction $U$ \cite{WBS+98}. 
Furthermore, we did not find a realistic 
temperature-driven RT within this approach. 

Thus we require an approximation scheme which is clearly beyond the
Hartree-Fock solution and takes into account 
electron correlations more reasonably. 
On the other hand, it must be simple enough to allow for an extended
study of magnetic phase transitions in thin films.
For this purpose we apply the spectral-density approach
(SDA)\cite{NB89,HN97a} which is motivated by the rigorous  analysis 
of Harris and Lange\cite{HL67} in the limit 
of strong Coulomb interaction ($U/t\gg1$). 
The SDA can be interpreted as an 
extension of their $t/U$-perturbation theory\cite{HL67,EO94} in a natural
way to intermediate coupling strengths and finite temperatures and 
has been  discussed in detail for various
three-dimensional \cite{NB89,HN97a,HN97b}  as well as 
infinite-dimensional\cite{HN97b,PHWN98} lattices. At least
qualitatively, it leads to rather convincing results concerning the magnetic 
properties of the Hubbard model.
A similar approach has been applied to a multiband Hubbard model with
surprisingly accurate results for the  magnetic key-quantities of the 
prototype band ferromagnets Fe, Co, Ni\cite{NBDF89,NVF95,VN96}.
Recently, a generalization of the SDA has been proposed to deal with the
modifications due to reduced translational symmetry\cite{PN96,PN97b}.
In the following we give only a brief derivation of the SDA solution and refer
the reader to previous papers for a detailed
discussion\cite{NB89,HN97a,PN96,PN97b}.

The basic quantity to be calculated is the retarded single-electron 
Green function 
$G_{ij\sigma}^{\alpha\beta}(E)=\langle\langle c_{i\alpha\sigma};
c_{j\beta\sigma}^{\dagger}\rangle\rangle_E$. 
From $G_{ij\sigma}^{\alpha\beta}(E)$ we obtain
all relevant information on the system. 
Its diagonal elements, for example, determine the 
spin- and layer-dependent quasiparticle density of states (QDOS)
$\rho_{\alpha\sigma}(E)=-\frac{1}{\pi}
\textrm{Im}G_{ii\sigma}^{\alpha\alpha}(E-\mu)$.
The equation of motion for the single-electron Green function reads:
\begin{equation} \label{eq_motion}
\sum_{l\gamma}\left[(E+\mu)\delta_{il}^{\alpha\gamma}-T_{il}^{\alpha\gamma} 
          -\Sigma_{il\sigma}^{\alpha\gamma}(E)\right]
          G_{lj\sigma}^{\gamma\beta}(E) =\hbar\delta_{ij}^{\alpha\beta}.
\end{equation}
Here we  have introduced the electronic self-energy
$\Sigma_{ij\sigma}^{\alpha\beta}(E)$ which incorporates all effects of electron
correlations.  
We adopt the local approximation for the self-energy which has been tested
recently for the case of reduced translational symmetry\cite{PN97c}.
If we assume translational invariance within each layer of the film we have
$\Sigma_{ij\sigma}^{\alpha\beta}(E)=\delta_{ij}^{\alpha\beta}
\Sigma^{\alpha}_{\sigma}(E)$. 
After Fourier transformation with respect to the 
two-dimensional Bravais lattice 
the equation of motion (\ref{eq_motion}) is formally solved by matrix
inversion.

The decisive step is to find a reasonable approximation for the
self-energy $\Sigma^{\alpha}_{\sigma}(E)$. 
Guided by the exactly solvable atomic limit of vanishing hopping ($t=0$)  
and  by the findings of Harris and Lange in the strong-coupling 
limit ($U/t\gg 1$), 
a one-pole ansatz for the self-energy $\Sigma^{\alpha}_{\sigma}(E)$ can be
motivated \cite{PN96}.
The free parameter of this ansatz are fixed  by exploiting the equality
between two alternative but exact representations
for the moments of the 
layer-dependent quasiparticle density of states:
\begin{equation}
\frac{1}{\hbar}\int dE (E-\mu)^{m}\rho_{\alpha\sigma}(E)=
\bigg\langle\Big[(i\hbar\frac{\partial}{\partial t})^m
c_{i\alpha\sigma}(t),c_{i\alpha\sigma}^{\dagger}(t')
\Big]_{+}\bigg\rangle_{t=t'} .
\end{equation}
Here, $[\dots,\dots]_{+}$ denotes the anticommutator.
It can be shown by comparing various approximation schemes \cite{PHWN98}
that an inclusion of the first four moments of the QDOS  ($m=0-3$)
is vital for 
a proper description of ferromagnetism  in the Hubbard model. 
Further, the inclusion of the first four moments
represents a necessary condition to be consistent
with the $t/U$-perturbation theory\cite{PHWN98}.
The Hartree-Fock approximation 
recovers the first two 
moments ($m=0,1$) only,  while the so-called Hubbard-I solution\cite{Hub63}
reproduces the third moment ($m=2$) as well, 
but is well known to be hardly able to describe ferromagnetism.
Taking into account the first four  moments to fix the free parameters
of our ansatz we end up with the SDA solution which is characterized by the 
following self-energy: 
\begin{equation}  \label{sig_sda}
     \Sigma_{\sigma}^{\alpha}(E) =
      U n_{\alpha-\sigma}
      \frac{E+\mu-B_{\alpha-\sigma}}{E+\mu-B_{\alpha-\sigma}-
	U\left(1- n_{\alpha-\sigma}\right)} 
\end{equation}
The self-energy  depends on the spin-dependent occupation numbers
$n_{\alpha\sigma}=\langle c_{i\alpha\sigma}^{\dagger}c_{i\alpha\sigma}\rangle$ 
as well as on the so-called 
bandshift $B_{\alpha\sigma}$ which consists of higher correlation functions:
\begin{equation}
   B_{\alpha\sigma}= T_{ii}^{\alpha\alpha}+
   \frac{1}{n_{\alpha\sigma}(1-n_{\alpha\sigma})}
	\sum_{j,\beta}^{j\beta\neq i\alpha} 
	T_{ij}^{\alpha\beta}\langle c_{i\alpha\sigma}^{\dagger} 
	c_{j\beta\sigma}(2n_{i\alpha-\sigma}-1)\rangle.
\end{equation} 
A possible spin dependence of $B_{\alpha\sigma}$ opens up the
way to ferromagnetic solutions\cite{NB89,HN97a}. 
Ferromagnetic order is indicated by a spin-asymmetry in the occupation numbers
$n_{\alpha\uparrow}\neq n_{\alpha\downarrow}$, and the layer-dependent 
magnetization
is  given by $m_{\alpha}=n_{\alpha\uparrow} - n_{\alpha\downarrow}$.  

The band occupations $n_{\alpha\sigma}$ are given by
\begin{equation}
   n_{\alpha\sigma}
   =\int\limits_{-\infty}^{+\infty} dE 
   f_{-}(E)\rho_{\alpha\sigma}(E),\label{nas}
\end{equation}  
where  $f_{-}(E)$ is the Fermi function. 
The mean band occupation $n$ is defined as
$n=\frac{1}{d}\sum_{\alpha\sigma}n_{\alpha\sigma}$.
Although $B_{\alpha-\sigma}$  consists of higher correlation functions 
it can by expressed
exactly \cite{NB89,HN97a} via $\rho_{\alpha\sigma}(E)$ and
$\Sigma_{\sigma}^{\alpha}(E)$:
\begin{eqnarray}
   B_{\alpha\sigma}&=&T_{ii}^{\alpha\alpha}+
   \frac{1}{n_{\alpha\sigma}(1-n_{\alpha\sigma})}
   \frac{1}{\hbar}\int\limits_{-\infty}^{+\infty} dE 
   f_{-}(E)\times\nonumber\\
   &&\hspace*{-4ex}\left(\frac{2}{U}\Sigma_{\sigma}^{\alpha}(E-\mu)-1\right)
   [E-\Sigma_{\sigma}^{\alpha}(E-\mu)-T_{ii}^{\alpha\alpha}]
   \rho_{\alpha\sigma}(E).
\label{bas}
\end{eqnarray}
Equations (\ref{eq_motion}), (\ref{sig_sda}), 
(\ref{nas}) and (\ref{bas}) build a
closed set of equations which can be solved self-consistently.

Surely a major short-coming of the SDA is the fact that quasiparticle damping 
is neglected completely. Recently a modified alloy analogy (MAA) has been
proposed \cite{HN96,NH98} which is also based on the exact results of the
$t/U$-perturbation theory but is capable of describing quasiparticle damping
effects as well. For bulk systems it has been found  that the magnetic 
region in the phase diagram is
significantly reduced by inclusion of damping effects. On the other hand,
the qualitative behavior of the magnetic solutions 
is very similar to the SDA. An application of the MAA
to thin film systems is in preparation\cite{HN98b}.

%
%
%
%
%
%
%
%
%

\section{Results and Discussion}\label{sec_res}

The numerical evaluations have been done for an fcc(100) film geometry. 
In this configuration each lattice site has four nearest neighbors 
within the same layer and four nearest neighbors in each of the respective 
adjacent layers. 
We consider uniform hopping $T_{ij}^{\alpha\beta}=-t$ between nearest neighbor
sites ${\bf R}_{i\alpha}$, ${\bf R}_{j\beta}$ only. Energy 
and temperature
units are chosen such that $t=1$. The on-site hopping integral is set to
$T_{ii}^{\alpha\alpha}=0$. 
Further, we keep the on-site Coulomb interaction fixed at $U=48$ 
which is three times the band width of the three-dimensional               
fcc-lattice and clearly refers to the strong-coupling regime.

\begin{figure}
	\centerline{\psfig{figure=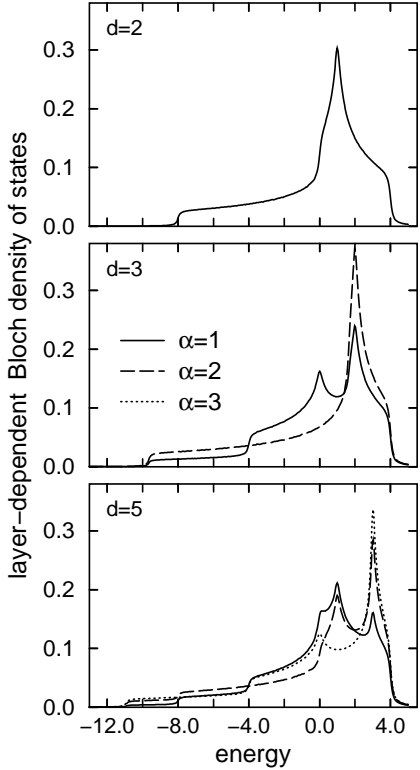,width=55mm,angle=0}}
	\vspace*{0.5cm}
	\caption{Layer-dependent Bloch density of states 
	        $\rho_{\alpha}^{(0)}(E)=\rho_{\alpha\sigma}(E)|_{U=0}$
	        of a two, three
                and five layer fcc(100) film. $\alpha$ denotes the different
                layers of the film  where $\alpha=1$ corresponds to 
                the surface-layer. 
                The nearest-neighbor hopping is set to $t=1$.}
	\label{fig_1}
\end{figure}

Let us consider the isotropic Hubbard-film first.
There are three model parameters left to vary, 
the temperature $T$, the thickness $d$
and the band occupation $(0\le n\le 2$). Except for the last part of the
discussion we will keep the band occupation fixed at the  value
$n=1.4$ and focus exclusively on the temperature 
and thickness dependence of the
magnetic properties.

In Fig.~\ref{fig_1} the  layer-dependent density of states of the 
non-interacting system 
$\rho_{\alpha}^{(0)}(E)=\rho_{\alpha\sigma}(E)|_{U=0}$ 
($\equiv$~``Bloch density of states'' (BDOS))
is plotted 
for a two, three and five layer film with an  fcc(100) geometry.
The BDOS is strongly asymmetric
and shows a considerable layer-dependence for $d\ge3$. 
Considering the moments
$\Delta_{\alpha}^{(n)}=\int dE (E-T_{ii}^{\alpha\alpha})^n
\rho_{\alpha}^{(0)}(E)$   of the BDOS yields 
that the variance $\Delta_{\alpha}^{(2)}$ as well as the skewness 
$\Delta_{\alpha}^{(3)}$ 
are reduced at the surface layer compared to the inner layers due to the
reduced coordination number at the surface
($\Delta_1^{(2)}=8$, $\Delta_{\alpha'}^{(2)}=12$, $\Delta_1^{(3)}=-24$, 
$\Delta_{\alpha'}^{(3)}=-48$ for
$\alpha'=2,\dots,d-1$).

\begin{figure}
	\centerline{\psfig{figure=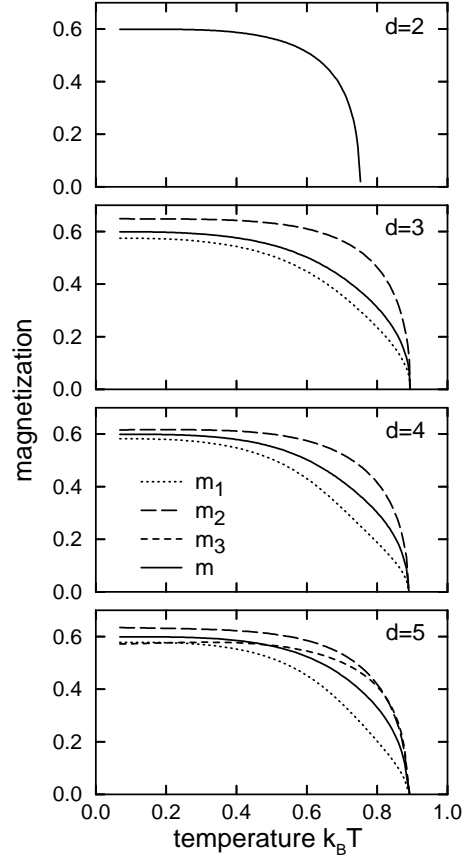,width=60mm,angle=0}}
	\vspace*{0.5cm}
	\caption{Layer magnetization 
	$m_\alpha$ and mean magnetization
	$m=\frac{1}{d}\sum_\alpha m_\alpha$   
	as a function of temperature for different number of
	layers $d=2,3,4,5$. 
	Further parameters: $n=1.4$, $U=48$.}
	\label{fig_2}
\end{figure}

In Fig.~\ref{fig_2} the layer magnetizations $m_\alpha$ as well as 
the mean magnetization $m=\frac{1}{d}\sum_\alpha m_\alpha$  are shown as a
function of temperature $T$. 
While symmetry requires the double layer film  to be 
uniformly magnetized,
the magnetization shows a   strong layer dependence for $d\ge3$.
The magnetization curves of the inner layers (and for the double layer)
show the usual Brillouin-type behavior. 
The trend of the surface layer magnetization 
($d\ge 3$), however,  is rather different. Note that $m_1$ 
depends almost linearly on   temperature in the range  
$T/T_C=0.7-0.9$ and for thicknesses $d\ge 4$. 
Compared to the inner layers, the surface magnetization  
decreases significantly 
faster as a function of temperature, tending to a reduced Curie temperature. 
However, due to the coupling between surface 
and inner layers  which is induced by the electron-hopping, 
this effect is delayed and a unique Curie-temperature for the whole film 
is found. 
The Curie temperature $T_C$  increases as a function of the film 
thickness $d$ 
and saturates already for film-thicknesses around $d=3-5$  to the
corresponding bulk value. 
A similar behavior was found for a  bcc(110) film geometry\cite{PN97b}.

\begin{figure}
	\centerline{\psfig{figure=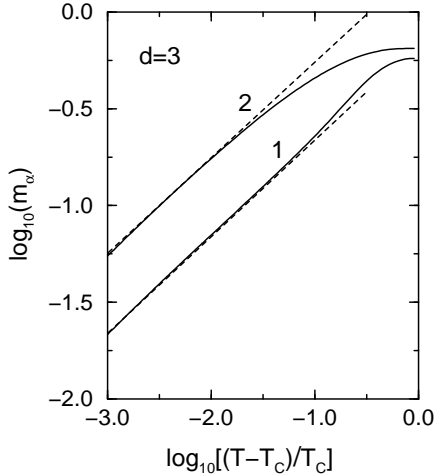,width=60mm,angle=0}}
	\vspace*{0.5cm}
	\caption{Temperature dependence of the 
	layer magnetization  $m_\alpha$ for a three-layer film
	on a logarithmic scale in the critical temperature
	range. The critical exponent of the magnetization is given by
	$\beta=1/2$ for all layers (dashed lines).
	Further parameters: $n=1.4$, $U=48$.}\label{fig_3} 
\end{figure}

The critical exponent of the magnetization (Fig.~\ref{fig_3}) 
is found to be equal to the mean-field value $\beta=\frac{1}{2}$ 
for all thicknesses and all other  parameters  considered. 
This clearly reveals the
mean-field type of our approximation which is 
due to the local approximation for
the electronic self-energy.

\begin{figure}
	\centerline{\psfig{figure=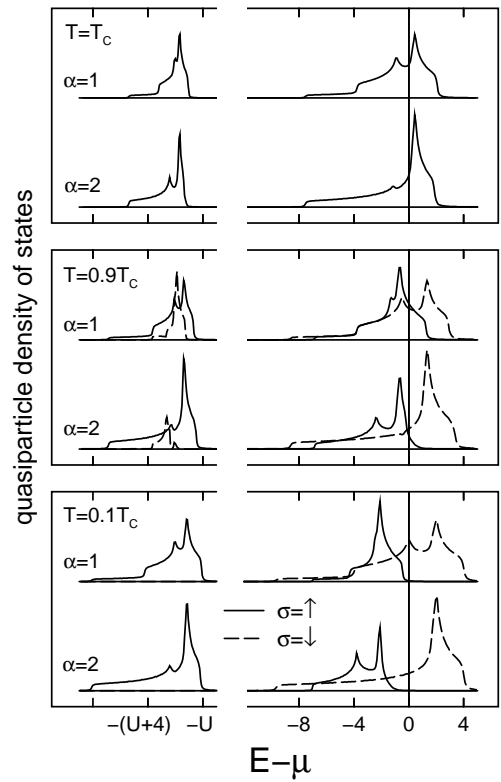,width=65mm,angle=0}}
	\vspace*{0.5cm}
	\caption{Spin-dependent quasiparticle densities of states for the outer
	and inner layer of a d=3 film.
	Solid lines:  majority spin direction ($\sigma=\uparrow$); 
	broken lines: minority spin direction ($\sigma=\downarrow$).
	Further parameters: $n=1.4$, $U=48$.}\label{fig_4}
\end{figure}

The layer-dependent quasiparticle density of states (QDOS) is shown in
Fig.~\ref{fig_4} for three temperatures $T/T_C=0.1,\,0.9,\,1.0$. 
Two kinds of splittings are observed in the spectrum.
Due to the strong Coulomb repulsion $U$ the 
spectrum splits into two quasiparticle
subbands (``Hubbard splitting'') which are separated by an  energy 
of the order $U$. In the lower subband the electron mainly hops over empty
sites, whereas in the upper subband it hops over sites which are already
occupied by another electron with opposite spin. 
The latter process requires an
interaction energy of the order of  $U$. 
The weights of the subbands scale with the probability of the realization 
of these two situations while the total weight of
the QDOS of each layer is normalized to 1. Therefore, 
the weights of the lower and upper
subbands 
are roughly given by $(1-n_{\alpha-\sigma})$ and $n_{\alpha-\sigma}$
respectively. This scaling becomes exact in the strong coupling limit 
($U/t\gg 1$).
Since the total band occupation ($n=1.4$) is above half-filling ($n=1$), the
chemical potential $\mu$ lies in the upper subband while the lower subband is
completely filled.

For temperatures below the Curie temperature $T_C$  an additional splitting
(``exchange-splitting'') in majority ($\sigma=\uparrow$) 
and minority ($\sigma=\downarrow$) spin
direction  occurs, leading to non-zero magnetization 
$m_\alpha=n_{\alpha\uparrow}-n_{\alpha\downarrow}$.
For $T=0$ the majority QDOS lies completely below the chemical potential,  
the system is fully polarized ($n_{\alpha\uparrow}=1$). Thus the
low-energy subband of the minority spin direction disappears and the
minority QDOS  is exactly given by the BDOS of the non-interacting system
(Fig.~\ref{fig_1}) due 
to vanishing correlation effects in the $\sigma=\downarrow$ channel. 
The reduced surface magnetization at $T=0$ (see Fig.~\ref{fig_2}) 
is, therefore, directly caused  by the layer-dependent  BDOS.

We like to stress that the spin-splitting does not depend on the size of the
Coulomb interaction $U$ as long as $U$ is chosen from  the  strong coupling
limit. Contrary to  Hartree-Fock theory  the spin splitting
saturates as a function of $U$ for values of about $2-3$ times 
the bandwidth of the
non-interacting system. 
The same holds for the Curie temperature $T_C$ \cite{HN97a}.

The temperature behavior of the QDOS is governed by two 
correlation effects (Fig.~\ref{fig_4}). 
As the temperature is increased, the spin-splitting between $\sigma=\uparrow$
and $\sigma=\downarrow$ spectrum decreases. This effect is accompanied by a
redistribution of spectral weight between the lower and upper subbands along
with a change of the widths of the subbands. 
For $T=0.9\,T_C$ one clearly sees that in the minority spectrum
weight has been transferred from the upper to lower  
subband which has reappeared due to
non-saturated magnetization at finite temperatures. In the  $\sigma=\uparrow$
spectrum the opposite behavior is found.
For all $T<T_C$ the spin splitting is significantly larger in the inner layer 
compared with the surface. 
At $T=T_C$ the exchange splitting has disappeared  whereas the
correlation-induced Hubbard splitting is still present.

Let us consider  the question why the surface magnetization shows a tendency
towards a reduced Curie temperature as seen in Fig.~\ref{fig_2}.
This effect can be understood by the above mentioned moment analysis of the 
BDOS (Fig.~\ref{fig_1}): 
From bulk systems it is known that an asymmetrically shaped BDOS is favorable
for the stability of ferromagnetism in the Hubbard model. 
In particular, the Curie temperatures
increase with increasing skewness of the BDOS \cite{WBS+98}.
The same trend shows up in the present 
film-system where the skewness of the BDOS is 
higher for the inner layers compared to
the surface-layer. 
Note that this argument is somewhat more delicate than in the case of 
Heisenberg films where the
reduced surface magnetization is directly caused by the reduced number of
interacting sites.

\begin{figure}
	\centerline{\psfig{figure=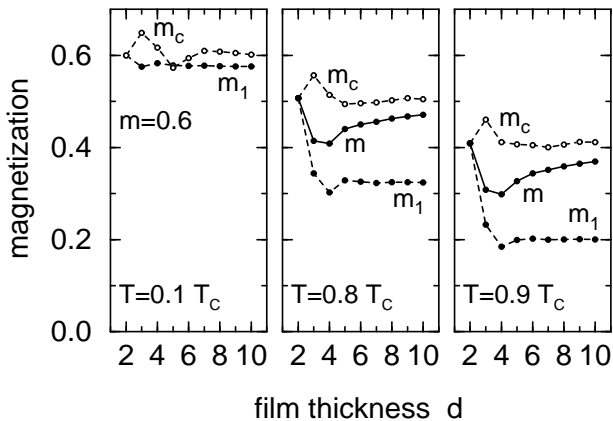,width=85mm,angle=270}}
	\vspace*{0.5cm}	
	\caption{Surface-layer magnetization $m_1$, central layer
	magnetization $m_c$ and mean 
	magnetization $m$ as a function of the film thickness $d$ and for three
	reduced temperatures $T/T_C=0.1,\,0.8,\,0.9$.
	Further parameters: $n=1.4$, $U=48$.}
	\label{fig_5} 
\end{figure}

In Fig.~\ref{fig_5} the difference between  surface magnetization $m_1$, 
central-layer magnetization $m_c$, and 
mean magnetization $m$ is analyzed in more detail. For all film thicknesses
 where this distinction is meaningful ($d\ge3$), the surface magnetization 
is reduced with respect to the mean magnetization. 
This holds not only for very
thin films ($d=3-5$, see also Fig.~\ref{fig_2}) where some oscillations  
are present that are caused by the finite film thickness, 
but also  extends to 
the limit $d\rightarrow\infty$ where the two surfaces  are well separated  
and do not interact.  The surface and central-layer 
magnetizations already stabilize for
thicknesses around $d=6$. 
Further, Fig.~\ref{fig_5} clearly shows that the reduction of 
$m_1$ drastically
increases for higher temperatures. For $T=0.9\,T_C$ 
the surface magnetization is
reduced to about half the size of the magnetization in the center of the film.

The charge transfer  due to differing layer occupations
$n_\alpha=n_{\alpha\uparrow}+n_{\alpha\downarrow}$ is found to be smaller 
than $\approx 0.03$ at $T=0$  and is almost negligible for finite temperatures.

We now like to focus on the 
magnetic anisotropy energy within the model system (\ref{ham_op}). 
The second-order anisotropy constant
$K_2=K_{2,\text{so}}+K_{2,\text{dip}}$ 
is calculated 
via Eq.~(\ref{f_anis_end}) which needs as an input the 
temperature-dependent layer magnetizations of the Hubbard film.
The dipole constants $\Phi_\alpha$ for an fcc(100) film geometry are found to 
be: 
$\Phi_0=0.7624\cdot 4\pi/3$, $\Phi_1=0.1206\cdot 4\pi/3$, 
$\Phi_2=-0.0020\cdot 4\pi/3$, and are set to zero for
$\alpha\ge 3$. 

To simulate both, surface and volume contribution of the spin-orbit induced
anisotropy, we choose the effective anisotropy field (\ref{anis_field}) 
in the surface-layer to be
different  from its value in the volume of the film:
\begin{equation}
\beta_\alpha=\left\{\begin{array}{ll}
              \beta_S\quad&\textrm{for }\alpha=1,d\\
              \beta_V     &\textrm{else.}
              \end{array}
              \right.
\end{equation}
In the perturbational approach only the ratio $\beta_\alpha/\omega_0$ 
is important. Thus we are left with only two parameters (fixed at $T=0$) 
to model the RT. 

\begin{figure}
	\centerline{\psfig{figure=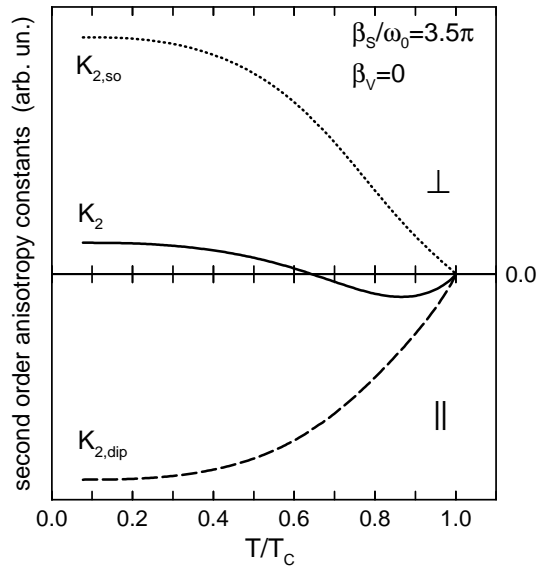,width=70mm,angle=0}}
	\vspace*{0.5cm}	
	\caption{Second-order anisotropy constants for a three-layer film in 
	the Fe-type configuration
	as a function of the reduced temperature $T/T_C$.
	Further parameters: $n=1.4$, $U=48$.}
	\label{fig_6}
\end{figure}

In principle these constants could be taken 
from experiment or from theoretical
ground-state  calculations. In our opinion, 
however, this would mean to somewhat
over-judge the underlying rather idealized model
system. We are mainly interested in the question whether 
a realistic temperature-driven RT is possible at 
all in the Hubbard film and by what mechanism it is
induced. 
Therefore, we choose the parameters 
$\beta_S$ and $\beta_V$  conveniently, guided
however by the experimental findings 
in the Fe-type and Ni-type scenarios described in the Introduction.

\begin{figure}
	\centerline{\psfig{figure=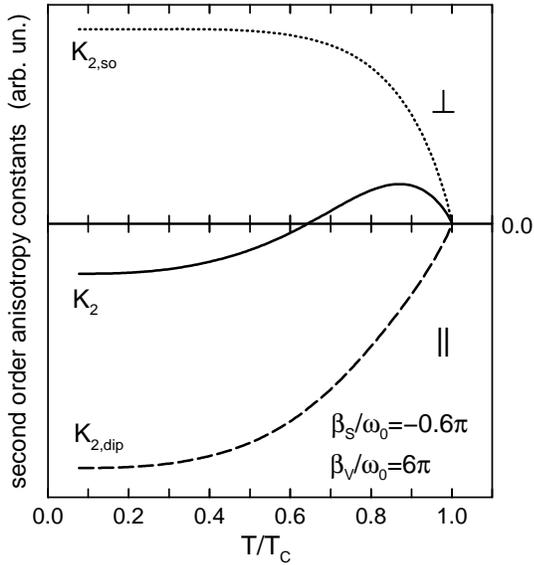,width=70mm,angle=0}}
	\vspace*{0.5cm}	
	\caption{Second-order anisotropy constants for a three-layer film in  
	the Ni-type configuration
	as a function of the reduced temperature $T/T_C$.
	Further parameters: $n=1.4$, $U=48$.}
	\label{fig_7}
\end{figure}

Fig.~\ref{fig_6} and Fig.~\ref{fig_7} show that for appropriate parameters
$\beta_S,\,\beta_V$ both types of temperature-driven RT can be found within a
three-layer film. The same is found for any film thickness $d\ge 3$.

In the Fe-type situation (Fig.~\ref{fig_6}) we 
consider a strong positive surface anisotropy
field $\beta_S/\omega_0=3.5\pi$ together with $\beta_V=0$.
At low temperatures the system is magnetized in out-of-plane direction. As the
temperature increases, $K_{2,\text{so}}$ decreases 
faster than $-K_{2,\text{dip}}$ because of
the strong reduction of the surface 
anisotropy and the magnetization switches to
an in-plane position. In principle this kind of RT is possible 
for all film thicknesses $d\ge 3$. For thicker films $\beta_S$ has to be
rescaled proportional to $d$ to compensate the increasing importance of the
dipole anisotropy.

The Ni-type  RT from in-plane to out-of-plane magnetization for a three layer
system is obtained by a positive volume anisotropy field $\beta_V=6\pi$ and 
a negative surface anisotropy $\beta_S=-0.6\pi$. At low temperatures the
dipole anisotropy as well as the negative surface anisotropy field lead to an
in-plane magnetization. For higher temperatures, however, where the surface
anisotropy becomes less important because of the reduced surface magnetization, 
the positive volume anisotropy field forces
the magnetization to switch to an out-of-plane direction. The ratio
$\beta_S/\beta_V$  determines for what thickness $d$ this type of 
temperature-driven RT 
is possible and scales like  $\beta_S/\beta_V\sim 1/d$ for thicker films. 

Note that for both types of RT the values of $\beta_S$ and $\beta_V$ are chosen
in such a way that the system is close to a thickness-driven RT.
In both cases the RT is mediated by the strong decrease of the
surface-layer magnetization compared to the inner layers as a function of
temperature. 

\begin{figure}
	\centerline{\psfig{figure=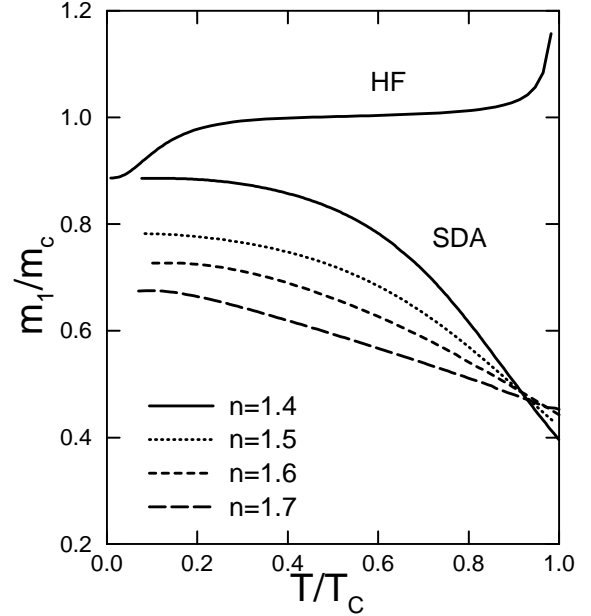,width=75mm,angle=270}}
	\vspace*{0.5cm}
	\caption{Ratio $m_1/m_c$ between surface- and center-layer magnetization
	as a function of the reduced temperature $T/T_C$ 
	for a three-layer film and different band occupations n. In addition
	$m_1/m_c$ calculated within Hartree-Fock theory is shown for
	$n=1.4$. Further parameter: $U=48$.}
	\label{fig_8}
\end{figure}

Finally, we consider the dependence
on the band occupation $n$. In
Fig.~\ref{fig_8} the ratio $m_1/m_c$ 
of a three-layer film is plotted as a function
of the reduced temperature $T/T_C$ for different band occupations 
$1.4\le n\le 1.7$. 
Above $n=1.78$ and below $n=1.0$ the fcc(100) Hubbard film 
does not have ferromagnetic solutions  whereas
between $n=1.0$ and $n=1.3$ there is a tendency towards first order phase
transition as a function of temperature\cite{HN97b}
and no realistic RT is possible.

The ratio $m_1/m_c$ (Fig.~\ref{fig_8}) 
decreases as a function of temperature for all band occupations.
For   $n=1.4$ which has been considered in
Fig.~\ref{fig_6} and Fig.~\ref{fig_7}, it changes 
from $m_1/m_c=0.9$ at $T=0$ to $m_1/m_c=0.4$ close to $T_C$. 
However, for higher 
band occupations this strong temperature-dependent  change in 
$m_1/m_c$ diminishes. From the  discussion above  it is clear that this is
unfavorable for the RT.
We can thus conclude from Fig.~\ref{fig_8} that the possibility of a
temperature-driven RT   sensitively depends on  the  band
occupation $n$. In addition we plotted in Fig.~\ref{fig_8} the
ratio $m_1/m_c$ calculated within Hartree-Fock theory. The result 
is completely different.
At low temperatures we find $m_1/m_c<1$ whereas $m_1/m_c>1$ close to
$T_C$. Note  that   the ratio $m_1/m_c$ is almost constant for
the wide temperature range $0.1<T/T_C<0.9$. 
Therefore, a realistic temperature-driven  
RT is excluded within the Hartree-Fock approximation.     
This holds for all parameters  that have been considered here.

%
%
%
%
%
%
%

\section{conclusion}\label{sec_con}
We have applied a generalization of the  
spectral-density approach (SDA) to thin Hubbard films. 
The SDA which reproduces the exact results of the $t/U$-perturbation theory in
the strong coupling limit, leads to rather convincing results concerning the
magnetic properties. The magnetic behavior 
of the itinerant-electron film can be 
microscopically understood by means of the temperature-dependent electronic
structure.

For an fcc(100) film geometry
the layer-dependent magnetizations have been discussed 
as a function of temperature
as well as  film thickness. The magnetization in the surface layer
is found to be reduced with respect to the inner layers for all thicknesses and
temperatures. By analyzing the layer-dependent 
QDOS this reduction can be
explained by the fact that in the free BDOS 
both variance and
skewness are diminished in the surface-layer compared to the inner layers.

The inclusion of the dipole interaction and an effective layer-dependent
anisotropy field  allows to study  the temperature-driven RT. The 
second-order anisotropy constants have been calculated within a perturbational
approach. For appropriate strengths of the 
surface and volume anisotropy fields
both types of RT, from out-of-plane to in-plane (Fe-type) and 
from in-plane to out-of-plane (Ni-type) magnetization are found. For the
Ni-type scenario the inclusion of a positive volume anisotropy is necessary. 
The RT in our itinerant model system is mediated by a strong reduction of the
surface magnetization with respect to the inner layers as a function of
temperature. Here a close similarity  to the model calculations within 
Heisenberg-type systems is apparent, 
despite the fact that these models completely
ignore the itinerant nature of the magnetic moments in the underlying
transition-metal samples. Contrary to Heisenberg films, 
the band occupation $n$ enters as an additional parameter 
within an itinerant-electron model. We find that
the possibility of a RT sensitively depends on the  band occupation. 
The fact that no realistic RT is possible within Hartree-Fock theory
clearly points out the importance of a reasonable treatment 
of electron correlation effects.

\acknowledgments{
This work has been done within the Sonderforschungsbereich 290 (``Metallische
d\"{u}nne Filme: Struktur, Magnetismus und elektronische Eigenschaften'') of the
Deutsche Forschungsgemeinschaft.}

\renewcommand{\baselinestretch}{1.0}
\Large

\end{document}